\documentclass[electronic]{vgtc}             

\ifpdf
  \pdfoutput=1\relax                   
  \usepackage{graphicx}                
  \DeclareGraphicsExtensions{.pdf,.png,.jpg,.jpeg} 
\else
  \usepackage{graphicx}                
  \DeclareGraphicsExtensions{.eps}     
\fi%

\graphicspath{{figures/}{pictures/}{images/}{./}} 

\usepackage{microtype}                 
\PassOptionsToPackage{warn}{textcomp}  
\usepackage{textcomp}                  
\usepackage{mathptmx}                  
\usepackage{times}                     
\usepackage{cite}                      
\usepackage{tabu}                      
\usepackage{booktabs}                  

\usepackage{hyperref}
\usepackage{xurl}
\usepackage{xcolor}

\usepackage{booktabs}
\usepackage{multirow}

\onlineid{0}

\vgtccategory{Research}

\vgtcinsertpkg

\title{Reducing the Human Factor in Virtual Reality Research\\to Increase Reproducibility and Replicability}

\author{Daniel Hepperle\thanks{e-mail: daniel.hepperle@h-ka.de}\\ %
      \parbox{2.4in}{\scriptsize \centering Karlsruhe University of Applied Science \& \\University of Hohenheim, Germany}
\and Tobias Dienlin\thanks{e-mail: tobias.dienlin@univie.ac.at}\\ %
     \scriptsize University of Vienna, Austria %
\and Matthias W\"olfel\thanks{e-mail: matthias.woelfel@h-ka.de}\\ %
     \parbox{2.4in}{\scriptsize \centering Karlsruhe University of Applied Science \& \\University of Hohenheim, Germany}}

\abstract{
The replication crisis is real, and awareness of its existence is growing across disciplines. We argue that research in human-computer interaction (HCI), and especially virtual reality (VR), is vulnerable to similar challenges due to many shared methodologies, theories, and incentive structures. For this reason, in this work, we transfer established solutions from other fields to address the lack of replicability and reproducibility in HCI and VR. We focus on reducing errors resulting from the so-called \textit{human factor} and adapt established solutions to the specific needs of VR research. In addition, we present a toolkit to support the setup, execution, and evaluation of VR research. Some of the features aim to reduce human errors and thus improve replicability and reproducibility. Finally, the identified chances are applied to a typical scientific process in VR.

} 

\CCScatlist{
\CCScatTwelve{Human-centered computing}{HCI theory}{concepts and models};
\CCScatTwelve{HCI design and evaluation methods}{user studies};

}

\begin{document}


\firstsection{Introduction}
\maketitle


The necessity for increased qualitative rigor in research (especially in regards to reproducibility and replicability) has spread around manifold disciplines ranging from (social) psychology~\cite{AAAS2015} over neuroscience~\cite{Button2013} to communication~\cite{Dienlin2020} and the social sciences in general~\cite{Camerer2018}. 
Even though the disciplines differ, they share a common set of methods (e.g., questionnaires), theories (e.g., technology acceptance model), and incentives (e.g., focus on publication quantity). Hence, we can assume that the problems originally causing the so-called \textit{replication crisis}---such as a lack of robustness and trustworthiness~\cite{AAAS2015}---are likely to exist in disciplines such as empirical 
 sciences and human-computer interaction (HCI) as well. 
Another reason could be that researchers from empirical computer science disciplines are not sensitive to these issues.  It seems that HCI and VR research lacks statistical power analyses and coherent approaches to increase reproducibility and replicability~\cite{Lanier2019, Cockburn2020}. 

Fortunately, however, disciplines such as computer sciences or HCI are per se qualified for systematic, structured procedures especially regarding approaches related to the open source community.  For example, using version control software~\cite{rodriguez2012distributed} such as GIT is a common practice in this area, and also for research purposes, it offers increased transparency and traceability. 
Hence, we argue that there are essential possibilities to increase research quality in terms of transparency and objectiveness in HCI by adapting established practices from neighboring disciplines, while also incorporating strengths typical of and unique to HCI. Many of the approaches that show high potential to reduce biases, heuristics, and other errors were already introduced by the open science community. 
Building on top of that, also in empirical computer sciences we see much potential for dedicated software-based approaches such as frameworks and toolkits that offer a sophisticated pipeline for research setup, test, and evaluation, flexible enough to be customized while powerful enough to provide the most common functions and methods to reduce human-induced errors.

In the following, we discuss these possibilities through the introduction and use of toolkits particularly designed for conducting empirical VR research. A good overview of such toolkits can be found in~\cite{woelfel21cyberworlds} which also briefly describes the \textit{Virtual Reality Scientific Toolkit} (VRSTK) introduced by the authors just recently. The main idea behind such toolkits is to support researchers to be more efficient by reducing the amount of redundant work that occurs on most VR research projects such as implementing questionnaires or data-logging. 
But, even more importantly, some of those toolkits are available open source which enables the community to constantly reevaluate and improve them, as well as to discuss improvements directly with the developers. When open source, most of the projects including their documentation are hosted in open repositories including a visible development history.

With this in mind, on the one hand, we see the possibility of using (predefined) computer instructions to create, modify, and capture all aspects of these virtual (research) worlds. On the other hand, the differences between the virtual and the real world give rise to various challenges---not only technical ones.  
Here we discuss the relationship between the traditional human factors that are crucial for reproducibility and replicability in research and the specific characteristics of VR that can support this. The main points discussed are: 

\begin{description}
	\item[Implementing Standards:] Frameworks or toolkits in computer science are an established way of reducing redundant work by providing a code base that offers a proven set of methods. This is discussed in two directions. First, providing an open source code base so that dedicated scientists are able to get a full overview of the code and also to participate in the development process. Second, providing scientifically proven solutions to common problems, such as best practice examples on how to implement questionnaires in VR to reduce the break in presence~\cite{Putze2020} or how to prepare recorded data for further analysis in statistical software such as R or similar.   
	\item[Process Documentation:] One of the reasons why research fails to replicate is missing or unclear documentation \cite{AAAS2015}. Broken down to its very essence, VR is a set of data and code in which all treatises can be logged and processed for further analysis.
	\item[Optimizing Generalizability:] In a traditional research environment, participants and investigators have predefined characteristics, such as size, ethnicity, or age, which cannot be altered. However, in VR, participants and investigators are represented by avatars that can freely be modified, liberating from given constraints and thus allowing for a setup that can be tailored to the individual user or experiment. 
	\item[Sharing Test Environments:] Since most of the parts in the VR research process are digital, the complete test environment can be saved and shared via online repositories such as github and gitlab\footnote{An overview of how to use version control software in order to increase reproducibility is given here by "theturingway" project~\cite{theturingway2019} }.
	``Barriers to effective data sharing and preservation are deeply rooted in the practices and culture
	of the research process as well as the researchers themselves. New mandates for data management plans from the National Science Foundation (NSF) and other federal agencies and world-wide attention to the need to share and preserve data could lead to changes.''~\cite{Tenopir2011}
	\item[Technical Limitations:] Although VR research equipment is generally less expensive than traditional laboratory facilities, researchers still need to invest in appropriate hardware, software, and content. Hardware is also generally developed in rapid iterations. This must be taken into account in terms of reproducibility and replicability.
	\end{description}

Taking these points into account, our goal in this paper is to provide interested researchers with an approach to improve their research process. Specifically, we aim to reduce errors resulting from human factors. We believe this can be achieved without much additional work, since the toolkits themselves have the goal of minimizing workload and the processes we present can be seen more as a by-product when using them.

\section{Theoretical Foundation}
``Everyone agrees that scientific studies should be reproducible and replicable. The problem is almost no one agrees upon what those terms mean.'' This vivid statement by Patil et al.~\cite{Patil2016} exemplifies the problem with the terminology between different research disciplines as well as within the same area of research. Widely accepted definitions of  \textit{replicablity} and \textit{reproducibility} are given by Asendorpf~\cite{Asendorpf2013} who have their origin in the field of psychology and are also applied in communication~\cite{Dienlin2020} and empirical computer sciences: 

\textit{Reproducibility} ``means that `Researcher B' [$\ldots$] obtains exactly the same results (e.g. statistics and parameter estimates) that were originally reported by `Researcher A' [$\ldots$]  from A’s data when following the same methodology.''

\textit{Replicability} ``means that the finding can be obtained with other random samples drawn from a multidimensional space that captures the most important facets of the research design.''

Another critical aspect in regards to the \textit{human error} is \textit{generalizablity}, which is defined as a finding that ``[$\ldots$]  does not depend on an originally unmeasured variable that has a systematic effect. [$\ldots$] \textit{Generalizability} requires replicability but extends the conditions to which the effect applies.''~\cite{Asendorpf2013, Dienlin2020}

With these definitions, we see two main points emerging which are prone to \textit{human error}. First, in case of \textit{reproducibility}, to achieve the exact same results as `Researcher A', it is necessary for `Researcher B' to: 

\begin{itemize}
	\item be able to gain the datasets used by `Researcher A', completely prepared and well documented for further use. 
	\item be able to use and work with data provided by `Researcher A'. This requires a comprehensive report about methodology as well as a clear understanding of the research question.
\end{itemize}
While there already exist several approaches on how to provide data beside the basic journal submission such as provided by the open science foundation\footnote{\url{https://osf.io/}}, it is still time consuming. Especially in regards to the \textit{publish or perish} concept, it means additional work that not everyone is given enough time for providing their (clean) datasets.

For \textit{replicability}, comprehensibility of the experimental procedure is of upmost importance. If possible, the best way to achieve this is to be present when the original research was carried out.
However, in many cases this is not possible and sometimes it's helpful if the authors do not know each other.
As a result, we believe it is necessary to create documentation that goes beyond the standard scope of a journal. 
This extended way of documenting the work should enable researchers to understand the different, sometimes vague, conclusions made in order to formulate hypothesis and reporting results. 

\textit{Generalizability} requires similar standards such as \textit{replicability}.
However, it adds even more need for documentation since it is rooted in the conflict between verbal expressions or qualitative theoretical claims and quantitative measure of it (i.e., construct validity~\cite{Cronbach1955}). 
Yarkoni~\cite{Yarkoni2020} for example emphasizes the fact that authors should ``disclose [$\ldots$] non-trivial effects (e.g. effects on stimuli, experimenter, research site, culture etc.) to the best of authors' abilities.''

\section{The Human Error}
A sole observation without any form of transcription or recording will lead to information loss.
Even with transcription, factors such as social cues, tonality, and non-verbal information get lost. 
This is not a new fact but has been observed already in the 80s, where almost 90\% of social science investigations were done using qualitative interviews~\cite{BRENNER1981139}. Since then (if not earlier), researchers suggest ubiquitous recording (both video and audio) for better comprehension\cite{briggs1986learning}. 
`` `human error' is not just about humans, it is about how features of people’s tools and tasks and working environment systematically influence human
performance.''~\cite{dekker2014} This can be interpreted in two directions. On the one hand, complicated system design increase errors. On the other hand, tools help reduce human-induced errors. 
Therefore, an interface easy to understand is necessary to reach as many people as possible, while also reducing the above mentioned kind of errors and challenges.

Even though there is a shift from qualitative approaches to quantitative research methods such as questionnaires---aiming to reduce subjective interpretation by making it quantifiable for example via Likert-scale ratings---these approaches are still prone to human error, for example with regard to inconsistencies in test-statistics~\cite{Nuijten2015} and refusal of data sharing~\cite{Tenopir2011}. 
This is especially concerning the high error rates and overall low transparency in the field of VR~\cite{Lanier2019}. 
In addition, only two out of 61 works provided supplementary material. Also, the rather low number of participants in the respective studies ($median=25$) implies that studies tend to be underpowered. To illustrate, chances of detecting a small to medium ($d=0.2; d=0.3$) effect with the given median participant number is between 17\% to 28\%~\cite{Lanier2019}.  
Inter-rater reliability (IRR)---which basically is a quantization of the mismatch between the different coders---often is not fully reported and the analysis of such is misinterpreted so that other researchers cannot address how the IRR affects the conducted evaluation~\cite{Hallgren2012}.

As pointed out earlier, several of these issues based on human error can be overcome by using structured approaches such as those offered by computer programs. 
The recent development of various toolkits to support the research process, in general, have the potential to counteract these sources of error in various ways.
 In the following sections, we will discuss these possibilities based on typical research steps in VR research.

\section{Toolkits for Virtual Reality Research}

Recently, a number of toolkits were introduced that aim to support research in VR~\cite{woelfel21cyberworlds}. Most of those toolkits share the goal to reduce redundancies by supporting recurring research tasks such as data logging and data export.  
While this is also one of the goals for the VRSTK, there exist several other functions that help to increase the research process in general and thus are introduced here briefly.

The VRSTK originated as a by-product of our ongoing research in the Intelligent Interaction \& Immersive Experience Lab at the Karlsruhe University of Applied Science by reducing redundancies in the context of data logging and data export. 
Over time, while learning more about the specific needs and chances of VR research, the set of functions implemented increased, and thus, we decided to take the project to another level by rethinking the whole process of (easy) implementation, standardization, and collaboration. 

As of now, the toolkit comes with several predefined scenes that provide easy access to technologies such as 
\begin{itemize}
\item basic interaction forms, 
\item questionnaire implementation, 
\item eye \& gaze tracking, or 
\item pose detection. 
\end{itemize}
Everything happening within the virtual test environment can be recorded in real-time for a later replay, augmented analysis, and export to dedicated statistic software such as R. 
The toolkit is implemented within the game engine Unity\footnote{\url{https://unity3d.com/}}.
It is provided as open-source software on github\footnote{\url{https://github.com/ixperience-lab/VRSTK/}}. 

Currently, next to a multiplayer setup for remote testing, the capturing of brain activities via EEG is implemented. Another important aspect of the VRSTK is the investigation and representation of humans which seems to be particularly challenging in VR in contrast to other media; see e.g., the uncanny valley effect~\cite{Hepperle2021}. For a full feature overview of the VRSTK see~\cite{woelfel21cyberworlds}. 

\section{Current Computational and Statistical Approaches to Reduce the Human Factor}

``I would hazard a guess that if we examined the combined effects of the numerous sources of bias that operate in various sociological and psychological studies we would find that they account for considerably more of the variance in the dependent variables of interest than do the major independent variables.''~\cite{Phillips73abandoningmethod}.

As stated before, also VR research lacks approaches to ensure replicability and reproducibility, which lowers trust in and also the applicability of those results~\cite{Wingen2019}.
However, in writing this manuscript for a dedicated workshop on replicability in extended reality, even though it is the first of its kind, we see an increased awareness, similar to what has happened in other disciplines before. 
Therefore, in what follows we look at valuable practices introduced in other disciplines that help to increase overall research quality such as the open science approach. We then adapt and extend them to the specific needs of VR research.

Dienlin et al.~\cite{Dienlin2020} suggested a 7-step open science agenda in the field of communication of which most points can be applied to many other research disciplines as well. 
The seven steps are as follows:
\begin{enumerate}
\item publish materials, data, and code
\item preregister studies and submit registered reports
\item conduct replications
\item collaborate
\item foster open science skills
\item implement transparency and openness promotion guidelines
\item incentivize open science practices
\end{enumerate}
Below, we discuss how we address the points 1, 3 and 4 with our toolkit.

Next to proposals on how to create more trust and reliability in research work regarding open science, there are calls to improve the way how results are reported. 
For example, Dragicevic~\cite{Dragicevic2016} promotes the use of confidence intervals instead of p-values for the field of computer science, aiming to provide a better interpretation of the results.

In addition to these specific approaches, web-based tools such as ``parsif.al''\footnote{\url{https://parsif.al}} or ``asreview.nl''\footnote{\url{https://asreview.nl/}} emerge to support the standardization approach for systematic review. The ``statcheck'' web interface\footnote{\url{http://statcheck.io/}} offers automated support in detecting errors in statistical reporting by simply uploading the PDF file.  

In what follows we apply those points to VR research, adding points where we see potential for such toolkits to offer a valuable contribution.

\begin{figure*}[h]
	\includegraphics[width=\textwidth]{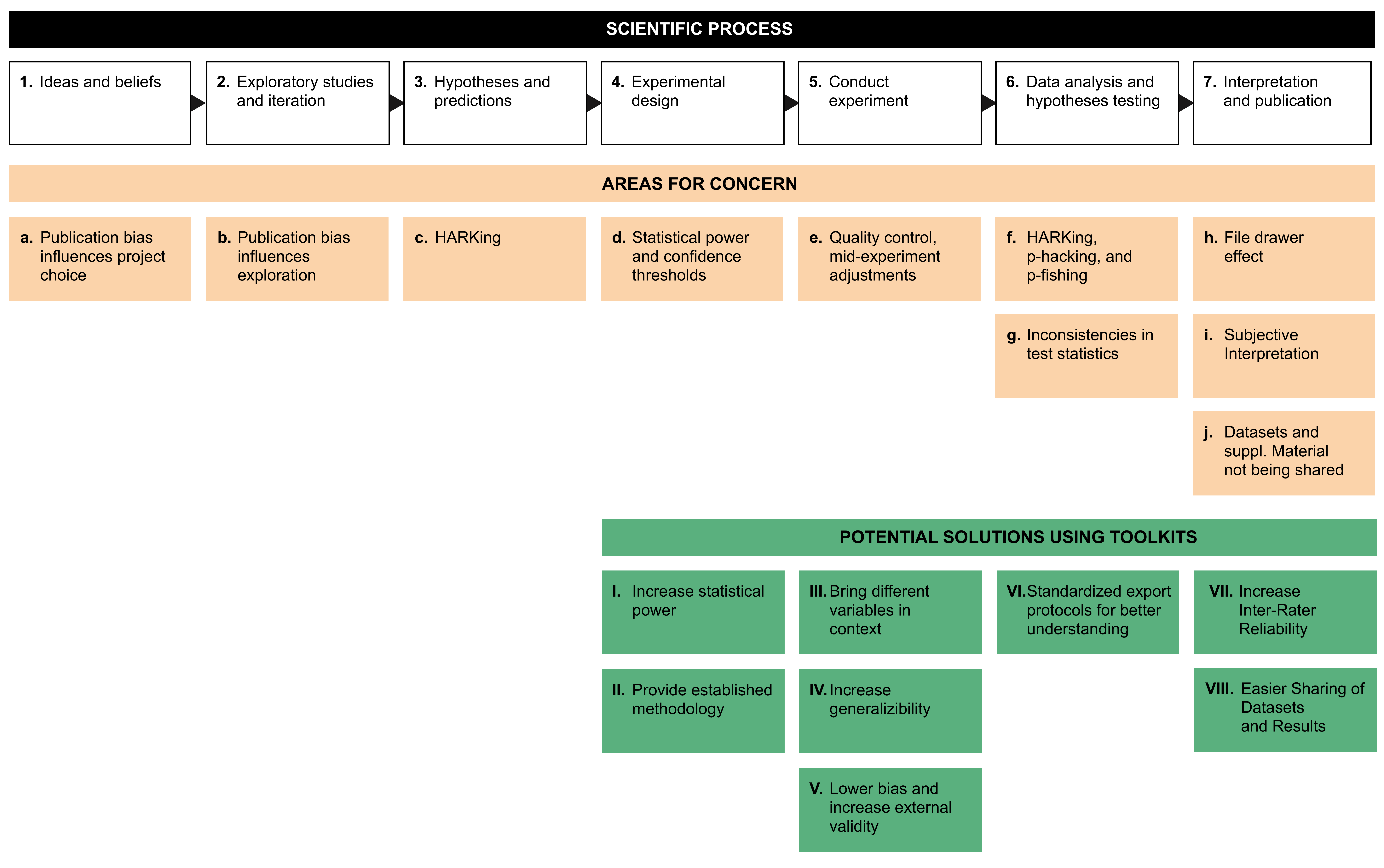} 

	\caption{ Stages of a typical experimental process (top) including typical areas of concern (mid) and potential solutions using toolkits (bottom). Originally introduced by \cite{Gundersen2018} and adapted from \cite{Cockburn2020}}
		\label{fig:scientific-process2}
\end{figure*}

\section{Steps to Improve Virtual Reality Study Procedures}

Most of the individual steps within a VR research process are similar to those from related disciplines, although some differ slightly. Looking at Fig.~\ref{fig:scientific-process2}, which shows a common research process, we identify \textit{step 4, 5, 6, and 7} for which mentioned toolkits are especially helpful.
Given that in quantitative research objectivity is the upper limit for reliability, one important goal is to eliminate subjective influence wherever possible. 

In the following, we discuss how mentioned problems concerning objectivity can be overcome or lowered by functions as offered by the use of toolkits. We will discuss the options based on 4 steps about the features currently provided---or planned to be implemented---in the VRSTK while they might also apply to other toolkits.\newline

\textbf{Step 4 -- Experimental Design} in which statistical power is a concern can be addressed by the VRSTK such that it facilitates sharing the whole research project with other researchers to increase the number as well as diversity of participants. 
Also, the option to apply a (remote) multiplayer setup, so that participants can join from either their living room with their technical setup or from a lab nearby, can increase participation and thereby statistical power (see Point I. in Fig.~\ref{fig:scientific-process2}).
Since it is always a critical factor for reproducible research to have enough participants, easy access to participation can increase participant number and therefore power. 

Another helpful aspect of formal toolkits is that they provide established methods critical for VR research (see Point II. in Fig.~\ref{fig:scientific-process2}).
This includes in-VR-questionnaires to keep up presence as well as other approaches to cope with fatigue, motion sickness, and other unwanted effects. For example, the ETRI (Korea's Electronics and Telecommunications Research Institute) has successfully trained an artificial intelligence that helps identify potential issues regarding motion sickness in VR applications\footnote{\url{https://disruptivetechasean.com/big\_news/technology-to-reduce-motion-sickness-in-vr-applications-developed-using-ai/}}.
 
Notably, within the VR environment, everything can be shared across the world.
However, this excludes aspects from reality in which the experiment takes place. 
Here aspects such as the general onboarding process (e.g. how to familiarize people with a headset), cultural background (e.g. preferences of the color temperature of the light), presumptions about the experiment, and other aspects such as outside temperature might lead to a potential bias---even though the virtual setup is identical. 
Hence, there still is a need for clear documentation besides sharing the sole codebase.\newline

\textbf{Step 5 -- Conduct Experiment } is where the VRSTK offers the greatest benefit. 
With the implementation of pre-defined procedures, (technical) generalizability can be increased. This is achieved by the following aspects of the VRSTK:

\begin{description}
	\item[Sensing Participants] A major feature of VR research is its ability to comprehend every single variable within the VR environment, creating a unique common context (see Point III. in Fig.~\ref{fig:scientific-process2}). 
	For example, it can be recorded in real-time where a participant looked at and for how long in a very exact manner using simple ray-casting. In addition, this can be annotated for example with simultaneous transcription of what currently is said in the experiment.		
	Even small differences in the implementation of the data processing of different sensor types can lead to a bias in results. With a predefined implementation in combination with comprehensive documentation, we hope for increased standardization and higher generalizability. Using a different microphone for example can lead to a different frequency pattern and therefore the algorithms’ interpretation of the speakers’ mood might differ 	(see Point IV. in Fig.~\ref{fig:scientific-process2}).  
	\item[Representation in VR] One of the most critical aspects of VR research is the representation of either the participant or the operator, which greatly influences the participant's perception of all interpersonal interactions within a virtual environment. 
	Representation can be divided into several parts. On the one hand, we have the outer appearance of the avatar which is defined by variables such as gender, height, ethnicity but also clothing. 
	Also depending on character style, the perception of those characters might be prone to the uncanny valley effect, which tends to be even more pronounced within VR environments than on 2D monitors for example~\cite{woelfel21cyberworlds}. 
	Without technical support by systems---such as Microsoft's Azure Kinect or HTC Vive Trackers combined with inverse kinematics---the transmission of the real participants' movement of extremities in VR is not given. 
	Therefore, the VRSTK offers a predefined scene that implements these transmission possibilities. Since these are rather complicated procedures, a common research or implementation standard would be helpful to foster generalizability, and possibly increase external validity. 
	Notably, one problem we see in this regard is varying technical setups. If one person for example lacks one of the technologies mentioned above, tracking is not possible and therefore the participant can either not take part or will create a large bias.
	Another aspect of representation within VR is the ability to alter the look of the person in real-time to the respective needs and to lower bias based on different ethnicity, height, age, or gender for example (see Point V. in Fig.~\ref{fig:scientific-process2}). 
	This can be taken even further in multi-user applications in which one participant can see other participants in an average height for the respective ethnicity while their perspective fits their real height and vice versa to foster external validity. 
	Also, movement can be adjusted in the same way to gain additional information on for example proxemics. Here, in each version of the virtual setup in the multi-user application, the movement (translation in the horizontal direction) of one participant is not transmitted to the second participant and therefore does not influence the personal proxemics' preference. 
\end{description}

\textbf{Step 6 -- Data Analysis and Hypotheses Testing}  is supported by the VRSTK in parts. An export of pre-selected data can be prepared for post-processing in statistical software such as R. Also custom scripts are offered for R to plot common data such as eye-tracking heat maps. 

Since research methods are manifold and it often differs from what the respective researcher aspires to implement based on their different needs, we did not implement further scripts about specific data analysis. 
The provided standardized export protocol though is essential for researchers to better comprehend the data.

Likewise, other researchers can more easily replicate or reproduce the findings (see Point VI. in Fig.~\ref{fig:scientific-process2}). 
Features such as API support to common questionnaire platforms such as soscisurvey\footnote{\url{https://soscisurvey.de}} are planned but not implemented yet.\newline

\textbf{Step 7 -- Interpretation and publication} Interpretation of results usually is prone to subjective influence. A way to overcome this issue is to work in groups of several persons to check if everyone would come to the same conclusion and measure the IRR for variances within the ratings (see Point VII. in Fig.~\ref{fig:scientific-process2}). 
Here, features such as the annotated replay function come in handy because each researcher can individually take a look at the complete research process, allowing computer-measured support in form of annotation such as current heart rate. 

Furthermore, the provision of supplementary material and data in the course of the publication means effort and expense which leads to the fact that this often does not happen. 
This timely effort can be reduced by toolkits due to the standardized exports and general sharing options~(see Point VIII. in Fig.~\ref{fig:scientific-process2}).

\section{Conclusion and Outlook}
In this paper, we showed that many of the findings identified and formulated in the context of the replication crisis likely apply also to research in VR. 
On the one hand, we see several indicators that show research in the HCI is similarly prone to questionable research practices that eventually lead to the lack of reproducibility, replicability, and trust in (this) research. On the other hand, venues such as the ``workshop on replication in extended reality'' and other approaches address, discuss and---most importantly---spread the word about the issues and raise overall awareness. 

Due to the basic systematics underlying computer programs, and therefore also VR systems, we argue that the field of HCI is particularly suitable for integrating parts of the processes in a typical experimental setup. 
Furthermore, we apply these findings to existing research toolkits, which were originally developed to help non-specialist or untrained users set up VR experiments and at the same time save redundant work. 
Our discussion showed that there is potential to further systematize the scientific discourse with given technology (in the sense of programs), to make the process and the insights more transparent and thus more open. 
Although the propagated precautions and the associated technology are still in their infancy, we recognize the potential to increase trust in research results in the field of VR. 
As part of the discussion, the VRSTK is presented. The VRSTK is made available as open-source and other scholars are explicitly invited to contribute by suggesting improvements or adding features. 

Since some points are predestined to increase reproducibility and replicability, in the foreseeable future we recommend developing a roadmap that takes into account additional features necessary to further implement processes and solutions as discussed in the course of the open science initiative. 

As of today, the interface of the tool is not yet sufficiently developed to guarantee error-free use for untrained users which, as discusses above can also lead to human-induced error in a typical research process. 
Also, it has not been conclusively clarified to what extent results can also be projected onto the real world. Approaches such as open hardware (e.g., the connection and design of open-source hardware components), which can be used to capture sensations from the real world and feed them to the virtual test setup could help improve this process. 

Here, too, we see a need for standardization, since the sensors used are noisy and error-prone. The VRSTK currently offers a serial interface for connecting such sensors, but this is no more sophisticated than reading out a simple data stream and documenting it.

Raising awareness in our opinion is a crucial first step for others to become part of projects such as ``theturingway'' or ``open science foundation'' to eventually create a critical mass to carry on the development of software tools such as the VRSTK or similar. 
While several of the open science suggestions imply a substantial rethinking of how research is conducted, the proposed toolkits are easy to implement and offer also benefits in case of time-reduction for creating research procedures in VR. 
In conclusion, next to the timely benefits mentioned above, implementing dedicated toolkits into the research process helps make research more robust, and credible. 
Since there has been increased attention in related disciplines to such challenges, there exist various solutions to counteract the problem.

Notably, data-sharing comes with concerns especially concerning privacy standards of the participants as well as of the cooperating funding partners from industry and other copyright reasons. 
Fortunately, there are already standards that might apply to these issue such as the DSGVO introduced by the German government in the `Bundesdatenschutzgesetz'\footnote{Translates to Federal Data Protection Act}. 
The principles of `Datensparsamkeit'\footnote{Translates to data minimization} mentioned in this context can also be applied to an open research process. 
Fundamental to this is the idea that only data which actually will be used later is saved and that the purpose of use must be presented in a simple and comprehensible manner. 
Thomson et al.~\cite{Thomson2005} for example propose the use of proxy data to make data available for secondary use. 

In conclusion, we hope the VRSTK helps readers increase their research quality by implementing standardized procedures that decrease errors resulting from human factors, which together should help reproduce, replicate and after all improve research in VR.
On top of that, ultimately, the question arises whether---or in how far---VR can be a valuable tool to conduct research that originally would have been conducted in the real  in order to reduce human error and to improve replicability and reproducibility.


\bibliographystyle{abbrv-doi}

\bibliography{template}

\begin{thebibliography}{10}

\bibitem{Asendorpf2013}
J.~B. Asendorpf, M.~Conner, F.~D. Fruyt, J.~D. Houwer, J.~J.~A. Denissen,
  K.~Fiedler, S.~Fiedler, D.~C. Funder, R.~Kliegl, B.~A. Nosek, M.~Perugini,
  B.~W. Roberts, M.~Schmitt, M.~A. G.~V. Aken, H.~Weber, and J.~M. Wicherts.
\newblock Recommendations for increasing replicability in psychology.
\newblock {\em European Journal of Personality}, 27(2):108--119, Mar. 2013.
  doi: {{%
10\hspace{.1pt}\discretionary{.}{%
}{.}\hspace{.4pt}1002\discretionary{/}{%
}{/}per\hspace{.1pt}\discretionary{.}{%
}{.}\hspace{.4pt}1919}}


\bibitem{BRENNER1981139}
M.~Brenner.
\newblock Problems in collecting social data: a review for the information
  researcher.
\newblock {\em Social Science Information Studies}, 1(3):139--151, 1981. doi:
  {{%
10\hspace{.1pt}\discretionary{.}{%
}{.}\hspace{.4pt}1016\discretionary{/}{%
}{/}0143\discretionary{%
}{-}{-}6236\discretionary{%
}{(}{(}81\discretionary{)}{%
}{)}90029\discretionary{%
}{-}{-}6}}


\bibitem{briggs1986learning}
C.~Briggs.
\newblock {\em Learning how to ask: a sociolinguistic appraisal of the role of
  the interview in social science research}.
\newblock Cambridge University Press, Cambridge Cambridgeshire New York, 1986.

\bibitem{Button2013}
K.~S. Button, J.~P.~A. Ioannidis, C.~Mokrysz, B.~A. Nosek, J.~Flint, E.~S.~J.
  Robinson, and M.~R. Munaf{\`{o}}.
\newblock Power failure: why small sample size undermines the reliability of
  neuroscience.
\newblock {\em Nature Reviews Neuroscience}, 14(5):365--376, Apr. 2013. doi:
  {{%
10\hspace{.1pt}\discretionary{.}{%
}{.}\hspace{.4pt}1038\discretionary{/}{%
}{/}nrn3475}}


\bibitem{Camerer2018}
C.~F. Camerer, A.~Dreber, F.~Holzmeister, T.-H. Ho, J.~Huber, M.~Johannesson,
  M.~Kirchler, G.~Nave, B.~A. Nosek, T.~Pfeiffer, A.~Altmejd, N.~Buttrick,
  T.~Chan, Y.~Chen, E.~Forsell, A.~Gampa, E.~Heikensten, L.~Hummer, T.~Imai,
  S.~Isaksson, D.~Manfredi, J.~Rose, E.-J. Wagenmakers, and H.~Wu.
\newblock Evaluating the replicability of social science experiments in nature
  and science between 2010 and 2015.
\newblock {\em Nature Human Behaviour}, 2(9):637--644, Aug. 2018.

\bibitem{Cockburn2020}
A.~Cockburn, P.~Dragicevic, L.~Besan{\c{c}}on, and C.~Gutwin.
\newblock Threats of a replication crisis in empirical computer science.
\newblock {\em Communications of the {ACM}}, 63(8):70--79, July 2020. doi: {{%
10\hspace{.1pt}\discretionary{.}{%
}{.}\hspace{.4pt}1145\discretionary{/}{%
}{/}3360311}}


\bibitem{theturingway2019}
T.~T.~W. Community, B.~Arnold, L.~Bowler, S.~Gibson, P.~Herterich, R.~Higman,
  A.~Krystalli, A.~Morley, M.~O'Reilly, and K.~Whitaker.
\newblock The turing way: A handbook for reproducible data science, 2019. doi:
  {{%
10\hspace{.1pt}\discretionary{.}{%
}{.}\hspace{.4pt}5281\discretionary{/}{%
}{/}ZENODO\hspace{.1pt}\discretionary{.}{%
}{.}\hspace{.4pt}3233853}}


\bibitem{Cronbach1955}
L.~J. Cronbach and P.~E. Meehl.
\newblock Construct validity in psychological tests.
\newblock {\em Psychological Bulletin}, 52(4):281--302, July 1955. doi: {{%
10\hspace{.1pt}\discretionary{.}{%
}{.}\hspace{.4pt}1037\discretionary{/}{%
}{/}h0040957}}


\bibitem{dekker2014}
S.~Dekker.
\newblock {\em The field guide to understanding ‘human error’}.
\newblock CRC press, 2014.

\bibitem{Dienlin2020}
T.~Dienlin, N.~Johannes, N.~D. Bowman, P.~K. Masur, S.~Engesser, A.~S.
  K\"{u}mpel, J.~Lukito, L.~M. Bier, R.~Zhang, B.~K. Johnson, R.~Huskey, F.~M.
  Schneider, J.~Breuer, D.~A. Parry, I.~Vermeulen, J.~T. Fisher, J.~Banks,
  R.~Weber, D.~A. Ellis, T.~Smits, J.~D. Ivory, S.~Trepte, B.~McEwan, E.~M.
  Rinke, G.~Neubaum, S.~Winter, C.~J. Carpenter, N.~Kr\"{a}mer, S.~Utz,
  J.~Unkel, X.~Wang, B.~I. Davidson, N.~Kim, A.~S. Won, E.~Domahidi, N.~A.
  Lewis, and C.~de~Vreese.
\newblock An agenda for open science in communication.
\newblock {\em Journal of Communication}, 71(1):1--26, Feb. 2020.

\bibitem{Dragicevic2016}
P.~Dragicevic.
\newblock {\em Fair Statistical Communication in HCI}, pp. 291--330.
\newblock Springer International Publishing, Cham, 2016. doi: {{%
10\hspace{.1pt}\discretionary{.}{%
}{.}\hspace{.4pt}1007\discretionary{/}{%
}{/}978\discretionary{%
}{-}{-}3\discretionary{%
}{-}{-}319\discretionary{%
}{-}{-}26633\discretionary{%
}{-}{-}6\_13}}


\bibitem{Gundersen2018}
O.~E. Gundersen and S.~Kjensmo.
\newblock State of the art: Reproducibility in artificial intelligence.
\newblock In {\em AAAI}, 2018.

\bibitem{Hallgren2012}
K.~A. Hallgren.
\newblock Computing inter-rater reliability for observational data: An overview
  and tutorial.
\newblock {\em Tutorials in Quantitative Methods for Psychology}, 8(1):23--34,
  Feb. 2012. doi: {{%
10\hspace{.1pt}\discretionary{.}{%
}{.}\hspace{.4pt}20982\discretionary{/}{%
}{/}tqmp\hspace{.1pt}\discretionary{.}{%
}{.}\hspace{.4pt}08\hspace{.1pt}\discretionary{.}{%
}{.}\hspace{.4pt}1\hspace{.1pt}\discretionary{.}{%
}{.}\hspace{.4pt}p023}}


\bibitem{Hepperle2021}
D.~Hepperle, C.~F. Purps, J.~Deuchler, and M.~W\"{o}lfel.
\newblock Aspects of visual avatar appearance: self-representation, display
  type, and uncanny valley.
\newblock {\em The Visual Computer}, June 2021. doi: {{%
10\hspace{.1pt}\discretionary{.}{%
}{.}\hspace{.4pt}1007\discretionary{/}{%
}{/}s00371\discretionary{%
}{-}{-}021\discretionary{%
}{-}{-}02151\discretionary{%
}{-}{-}0}}


\bibitem{Lanier2019}
M.~Lanier, T.~F. Waddell, M.~Elson, D.~J. Tamul, J.~D. Ivory, and
  A.~Przybylski.
\newblock Virtual reality check: Statistical power, reported results, and the
  validity of research on the psychology of virtual reality and immersive
  environments.
\newblock {\em Computers in Human Behavior}, 100:70--78, 2019. doi: {{%
10\hspace{.1pt}\discretionary{.}{%
}{.}\hspace{.4pt}1016\discretionary{/}{%
}{/}j\hspace{.1pt}\discretionary{.}{%
}{.}\hspace{.4pt}chb\hspace{.1pt}\discretionary{.}{%
}{.}\hspace{.4pt}2019\hspace{.1pt}\discretionary{.}{%
}{.}\hspace{.4pt}06\hspace{.1pt}\discretionary{.}{%
}{.}\hspace{.4pt}015}}


\bibitem{Nuijten2015}
M.~B. Nuijten, C.~H.~J. Hartgerink, M.~A. L.~M. van Assen, S.~Epskamp, and
  J.~M. Wicherts.
\newblock The prevalence of statistical reporting errors in psychology
  (1985{\textendash}2013).
\newblock {\em Behavior Research Methods}, 48(4):1205--1226, Oct. 2015. doi:
  {{%
10\hspace{.1pt}\discretionary{.}{%
}{.}\hspace{.4pt}3758\discretionary{/}{%
}{/}s13428\discretionary{%
}{-}{-}015\discretionary{%
}{-}{-}0664\discretionary{%
}{-}{-}2}}


\bibitem{AAAS2015}
{Open Science Collaboration}.
\newblock Estimating the reproducibility of psychological science.
\newblock {\em Science}, 349(6251), 2015. doi: {{%
10\hspace{.1pt}\discretionary{.}{%
}{.}\hspace{.4pt}1126\discretionary{/}{%
}{/}science\hspace{.1pt}\discretionary{.}{%
}{.}\hspace{.4pt}aac4716}}


\bibitem{Patil2016}
P.~Patil, R.~D. Peng, and J.~T. Leek.
\newblock A statistical definition for reproducibility and replicability.
\newblock {\em BioRxiv}, 2016.

\bibitem{Phillips73abandoningmethod}
B.~Phillips and J.~C. Gazet.
\newblock Abandoning method, 1973.

\bibitem{Putze2020}
S.~Putze, D.~Alexandrovsky, F.~Putze, S.~H\"{o}ffner, J.~D. Smeddinck, and
  R.~Malaka.
\newblock Breaking the experience: Effects of questionnaires in {VR} user
  studies.
\newblock In {\em Proceedings of the 2020 {CHI} Conference on Human Factors in
  Computing Systems}. {ACM}, Apr. 2020. doi: {{%
10\hspace{.1pt}\discretionary{.}{%
}{.}\hspace{.4pt}1145\discretionary{/}{%
}{/}3313831\hspace{.1pt}\discretionary{.}{%
}{.}\hspace{.4pt}3376144}}


\bibitem{rodriguez2012distributed}
C.~Rodr{\'\i}guez-Bustos and J.~Aponte.
\newblock How distributed version control systems impact open source software
  projects.
\newblock In {\em 2012 9th IEEE Working Conference on Mining Software
  Repositories (MSR)}, pp. 36--39. IEEE, 2012.

\bibitem{Tenopir2011}
C.~Tenopir, S.~Allard, K.~Douglass, A.~U. Aydinoglu, L.~Wu, E.~Read, M.~Manoff,
  and M.~Frame.
\newblock Data sharing by scientists: Practices and perceptions.
\newblock {\em {PLoS} {ONE}}, 6(6):e21101, June 2011. doi: {{%
10\hspace{.1pt}\discretionary{.}{%
}{.}\hspace{.4pt}1371\discretionary{/}{%
}{/}journal\hspace{.1pt}\discretionary{.}{%
}{.}\hspace{.4pt}pone\hspace{.1pt}\discretionary{.}{%
}{.}\hspace{.4pt}0021101}}


\bibitem{Thomson2005}
D.~Thomson, L.~Bzdel, K.~Golden-Biddle, T.~Reay, and C.~A. Estabrooks.
\newblock Central questions of anonymization: A case study of secondary use of
  qualitative data.
\newblock {\em Forum Qualitative Sozialforschung / Forum: Qualitative Social
  Research}, 6(1), Jan. 2005. doi: {{%
10\hspace{.1pt}\discretionary{.}{%
}{.}\hspace{.4pt}17169\discretionary{/}{%
}{/}fqs\discretionary{%
}{-}{-}6\hspace{.1pt}\discretionary{.}{%
}{.}\hspace{.4pt}1\hspace{.1pt}\discretionary{.}{%
}{.}\hspace{.4pt}511}}


\bibitem{Wingen2019}
T.~Wingen, J.~B. Berkessel, and B.~Englich.
\newblock No replication, no trust? how low replicability influences trust in
  psychology.
\newblock {\em Social Psychological and Personality Science}, 11(4):454--463,
  Oct. 2019. doi: {{%
10\hspace{.1pt}\discretionary{.}{%
}{.}\hspace{.4pt}1177\discretionary{/}{%
}{/}1948550619877412}}


\bibitem{woelfel21cyberworlds}
M.~W\"olfel, D.~Hepperle, C.~Purps, J.~Deuchler, and W.~Hettmann.
\newblock Entering a new dimension in virtual reality research: An overview of
  existing toolkits, their features, and challenges.
\newblock In {\em Proceedings of Cyberworlds}, 2021.

\bibitem{Yarkoni2020}
T.~Yarkoni.
\newblock The generalizability crisis.
\newblock {\em Behavioral and Brain Sciences}, pp. 1--37, Dec. 2020. doi: {{%
10\hspace{.1pt}\discretionary{.}{%
}{.}\hspace{.4pt}1017\discretionary{/}{%
}{/}s0140525x20001685}}


\end{thebibliography}
\end{document}